\documentstyle[12pt,epsf]{article}
\def\gmmu{\gamma _{\mu}}

\def\gmf{\gamma _{5}}

\def\ll{\langle }
\def\rl{ \rangle }

\newcommand{\beq}{\begin{equation}}
\newcommand{\eeq}{\end{equation}}
\newcommand{\bea}{\begin{eqnarray}}
\newcommand{\eea}{\end{eqnarray}}

\begin{document}
\renewcommand{\thefootnote}{\fnsymbol{footnote}}
                                        \begin{titlepage}
\begin{flushright}
TECHNION-PHYS-95-13-REV \\
hep-ph/9506264
\end{flushright}
\vskip1.8cm
\begin{center}
{\LARGE
Radiative decay $ B \rightarrow l \nu \gamma $ in the light cone QCD
approach
            \\ }
\vskip1.5cm
 {\Large G.~Eilam , I.~Halperin , R.R.~Mendel}$\footnote { On sabbatical
leave from Dept. of Applied Math., Univ. of Western Ontario, London,
Ontario, Canada } $
 \\
\vskip0.2cm
       Technion - Israel Institute of Technology   \\
       Department of Physics  \\
       Haifa, 32000,  Israel \\
{\small e-mail addresses: phr82ge@technion.technion.ac.il \\
                          higor@techunix.technion.ac.il  \\
                          mendel@uwo.ca }\\
\vskip1.5cm
{\Large Abstract:\\}
\parbox[t]{\textwidth}{
We calculate the rate for the decay $ B_{u}
\rightarrow l \nu \gamma $ using the light cone QCD sume rules. We find
$ Br( B_{u} \rightarrow l \nu \gamma) \simeq 2 \cdot 10^{-6} $. The results
are used to test the applicability of the constituent quark model
approximation to the same process.The latter estimate is proportional to
$ 1/m_{u}^{2} $, where $ m_{u} \simeq \bar{\Lambda}_{u} $ is the "constituent
quark mass", indicating that the process is of long distance type. We find
that the two approaches yield similar results for the total rate with the
choice $ m_{u} \simeq 480 \;MeV $. This indicates that the constituent quark
model may be used for estimates of the radiative "annihilation" contribution
to  this and other radiative decays. We point out that this decay may be
useful for the measurement of $ |V_{ub}|$. }

\vspace{1.0cm}
{\em submitted to Physics Letters B }
\end{center}
                                                \end{titlepage}

\section{Introduction}
 Recently there has been increasing interest in long distance
contributions to
inclusive \cite{BSS,EMM,DHT} and exclusive \cite{MS,BGW,ABBS,AES,EJM}
radiative D- and B-decays. One way of estimating the long distance part
of annihilation-type contributions is to use a simple constituent quark model
in which the photon (or gluon) is emitted from the light quark in the
initial state \cite{BSS,EMM,MS,ABBS,AES,EJM}. Thus for radiative
decays such as $ B \rightarrow l \nu \gamma $ or $ B \rightarrow \rho
\gamma $ one finds that the annihilation amplitude is proportional to
$ 1/m_{q} $, where $ q $ is the light quark in the decaying meson. This
indicates that it is indeed of the long distance type in an average
sense, i.e. it is equivalent to a sum of all possible hadronic intermediate
states \cite{BGW}. One expects that more detailed observables, such as the
photon spectrum in $ B \rightarrow l \nu \gamma $, will not be reproduced
in full detail by the quark model.

In a recent paper \cite{AES} the reaction $ B \rightarrow l \nu \gamma $
was discussed using the annihilation static quark diagram with the photon
emitted from the u-quark giving the largest contribution. The result of
this calculation exhibits the particular $ 1/m_{u}^{2} $ dependence (see Eq.
(25) below). One may expect that in the case at hand this constituent quark
mass $ m_{u} $
must be close to the "inertia parameter" $ \bar{\Lambda}_{u} $ of the
Heavy Quark Effective Theory, which is defined as $ \bar{\Lambda} =
\lim_{m_{Q} \rightarrow \infty} (M_{(\bar{Q}q)} - m_{Q} )$. Estimates of $
\bar{\Lambda}_{u} $ typically yield a somewhat larger value for this
quantity than the number
 $ m_{u} \simeq 350 \; MeV $, which holds for light hadrons (see e.g.
\cite{Neu}).

In this letter we suggest a different and largely model-independent method
to  calculate the
differential and total width of the decay $ B \rightarrow l \nu \gamma $.
Our approach is based on the use of the concept of duality in a
form suggested by the QCD sum rules method \cite{SVZ}. In this
approach an
amplitude of interest is extracted from the imaginary part of a suitably
defined Euclidean correlation function. More specifically,
we use here a version of the QCD sum rule technique known as the light cone
QCD sum rules. The method has been developed for light quark systems in
Ref. \cite{BBK},\cite{BF1}. It has proved to be a convenient
tool for a study of exclusive processes with the emission of a light
particle.
An Euclidean correlation function corresponding to such a process can be
calculated using the light final particle as a source of a
properly defined variable external field in which the correlation
function develops. It turns out that in this case different operators
contribute according to their twist rather than their dimension. Matrix
elements of non-local operators in the variable external field are identified
with the set of wave functions of increasing twist, and replace vacuum
expectation values of local operators which appear in the traditional sum
rules method.  The behavior of wave functions is severely
restricted by the (approximate) conformal invariance of QCD. More
details on the method and a list of references can be found in \cite{BH},
\cite{BBKR}, see also Sect. 2 and 3 below.

In our problem we deal with a set of the photon wave functions of
increasing twist. Photon matrix elements are defined via a sum of matrix
elements containing vector meson states (see sect.3 for more details).
Unlike perhaps more complicated
hadron (such as $ \pi $- or $ \rho $-) wave functions, which
are still debatable in the literature, the photon wave functions turn out
to be  rather simple and are given by their asymptotic expressions. Thus
theoretical uncertainties are minimal in this case. We
note in passing that our approach in not equivalent
to the direct use of the vector meson dominance applied in \cite{ABBS}
to compare with the quark model annihilation contribution
in the decay $ B \rightarrow \rho \gamma $. The paper is organized as
follows. In sect.2 we obtain the light cone sum rule for the process of
interest to the twist 2 accuracy (with partial account for twist 4
effects). In sect.3 we collect the necessary
information on the photon wave functions of twist 2. The final sum rule
and a numerical analysis are presented in sect.4. We find transition form
factors and calculate the differential and total decay width. The results
are then confronted with those of Ref. \cite{AES}. We find that the
rates obtained by these two methods are the same
for the values of the constituent quark mass $ m_{u} \simeq 480 \; MeV
$. This range is in reasonable agreement with independent estimates for $
\bar{\Lambda} $. These results  indicate that the
quark model approximation may be used for estimates of the radiative
"annihilation" contribution to this and other heavy meson decays, in
which the photon is dominantly emitted from the light quark, such as
$ \bar{B}^{0} \rightarrow D^{0 \star} \gamma \, , \, B^{\pm} \rightarrow
\rho^{\pm} \gamma \, , \, B_{s} \rightarrow \gamma \gamma $. Sect. 5
contains a summary and several concluding remarks. We finally comment on
the possible usefulness of this decay for the measurement of $ |V_{ub}| $
and discuss
some general problems related to the annihilation mechanism in exclusive
amplitudes.

 \section{Light cone sum rule}

The effective Hamiltonian for the decay of interest is
\beq
H_{eff} = \frac{G_{F}}{\sqrt{2}} V_{ub} \;  \bar{\nu} \gmmu (1- \gmf) l
\;
 \ll \gamma | \bar{u} \gmmu (1- \gmf) b | B \rl
\eeq
We have neglected here the helicity-suppressed photon emission from the
final lepton.
We use the following parametrization for the hadron transition form factors
\bea
\frac{1}{\sqrt{4 \pi \alpha}} \ll \gamma (q)| \bar{u} \Gamma_{\mu} b | B(p)
\rl &=&  \varepsilon_{\mu \nu
\rho \sigma} e_{\nu}^{*} p_{\rho} q_{\sigma}
\frac{g(Q^{2})}{m_{B}^{2}}   \\
    &+& i (e_{\mu}^{*}( pq) - (e^{*}p) q_{\mu}) \frac{f(Q^{2})}{m_{B}^{2}}
 \nonumber
\eea
(here $ \Gamma_{\mu} = \gmmu ( 1- \gmf) $ , $ Q = p - q $ and $
e_{\mu}^{*} $ stands for the polarization vector of the photon.)
Our aim is to calculate the transition form factors $ g \; , \; f $
 including their momentum dependence. To this end we use the light
cone QCD sum rules method. Technically, our method is very close to the
calculation of the semileptonic $ B \rightarrow \pi e \nu $
\cite{BKR} and radiative $ B \rightarrow \rho + \gamma $, $
B \rightarrow K^{\star} + \gamma $ \cite{ABS} form
factors. These calculations are based on using the light cone wave
functions of a light pseudoscalar ( $ \pi $) or vector ( $ \rho ,
K^{\star} $) meson, respectively. In our problem, the relevant objects are
the light cone photon wave functions, which will be discussed below.

We start with the correlation function
\bea
T_{\mu}(p,q) &=& i \int
d^{4}x \, e^{ipx} \ll \gamma(q) |
T \{ \bar{u}(x) \gmmu (1- \gmf) b(x) \bar{b}(0) i \gmf u(0) \} |0
\rl  \nonumber \\
                &=& \frac{m_{B}^{2}f_{B}}{m_{b}} \frac{1}{m_{B}^{2} -
(p+q)^{2}} \, \ll \gamma (q)|\bar{u} \gmmu (1- \gmf) b| B(p+q) \rl +
\ldots \\
                &=& \sqrt{4 \pi \alpha} \left\{ \varepsilon_{\mu \nu
\rho \sigma} e_{\nu}^{*} p_{\rho} q_{\sigma} T_{1} +
i (e_{\mu}^{*}( pq) - (e^{*}p) q_{\mu}) T_{2} \right\} \nonumber
\eea
where $ \ll B | \bar{b} i \gmf u | 0 \rl = m_{B}^{2} f_{B} /m_{b} $ .
The second line in the above formula selects the pole contribution
due to the B-meson pole in the corresponding dispersion integrals for the
scalar form factors $ g $ and $ f $. On the other hand, the
correlation function (3) can be calculated in QCD at large Euclidean
momenta $ ( p+ q)^{2} $.

In this case the virtuality of the heavy quark, which is of order $ m_{b}^{2}
- (p+q)^{2} $, is the large quantity. Thus, we can expand the heavy quark
propagator in powers of slowly varying fields residing in the photon,
which act as external fields on the propagating heavy quark. The
expansion in powers of an external field is also the expansion of the
propagator in powers of a deviation from the light cone $ x^{2} \simeq 0
$. The leading
contribution is obtained by using the free heavy quark propagator in the
correlation function (3). We obtain
\beq
T_{\mu}(p,q) =  \int \frac{d^{4}x d^{4}k}{(2 \pi)^{4} ( m_{b}^{2}
- k^{2})} e^{i(p-k)x} \langle \gamma (q) | \bar{u}(x)
\gmmu (1- \gmf) (k_{\mu} \gmmu + m_{b}) i  \gmf u(0) | 0 \rangle
\eeq
In this formula a
path-ordered gauge factor between the quark fields is implied, as
required by gauge invariance. In the particular case of the
Fock-Schwinger gauge $ x_{\mu} A_{\mu}(x) = 0 $ this
factor is equal to unity.

We see that the answer is expressed via the one-photon matrix element of
the gauge invariant non-local operator with a light-like separation $ x^{2}
\simeq 0 $. This matrix element defines a light cone photon wave function
(WF) in a way analogous to the definition of the light cone meson WF's
\cite{CZ}. We define  two-particle photon WF's as follows :
\bea
\ll \gamma (q) | \bar{u}(x) \sigma_{\mu \nu} u(0) |0 \rl &=&  i
\sqrt{4 \pi \alpha}  e_{u} \ll \bar{u} u \rl \int_{0}^{1} du \; e^{iuqx}
[ ( e_{\mu} q_{\nu} - e_{\nu} q_{\mu} )( \chi
\phi(u)  \nonumber \\
&+&  x^{2} g^{(1)}(u)) + \{ (qx)(e_{\mu} x_{\nu} - e_{\nu}
x_{\mu})
+ (ex) ( x_{\mu} q_{\nu} -  x_{\nu} q_{\mu})    \nonumber \\
&-&  x^{2} (e_{\mu} q_{\nu} - e_{\nu} q_{\mu}) \}
g^{(2)}(u) ]  \\
\ll \gamma (q) | \bar{u}(x) \gmmu \gmf u(0)|0 \rl &=&
\frac{1}{4} \sqrt{4 \pi \alpha}
\varepsilon_{\mu \nu \rho \sigma} e_{\nu} q_{\rho}x_{\sigma}f
\int_{0}^{1} du \; e^{iuqx} g_{\perp}(u)
 \eea
Here $ \phi(u)\, , \,  g_{\perp}(u) $ stand for the leading twist 2 photon
WF, while $ g^{(1)}$
and $ g^{(2)} $ are the two-particle WF's of twist 4 (see below for more
detail).
The above WF's describe a distribution in the fraction of the total momentum
carried by the quark $(u p_{z})$ and the anti-quark $ ((1-u)p_{z})$ in the
infinite-momentum frame $ p_{z} \rightarrow \infty $ (or, equivalently, a
distribution in the light cone
momentum $ P_{+} $). The dimensional constants $ \chi $ and $ f $
are  chosen in such a way that a function $ f(u) = ( \phi(u) \, , \,
g_{\perp}(u) ) $ is normalized to unity :
\beq
\int_{0}^{1} du \; f(u) = 1
\eeq
It is assumed that perturbative logs of $ x^{2} $ are resummed by
renormalization group methods and result in a dependence of the WF's on
the scale factor $ \mu^{2} \sim x^{-2} \sim m_{b}^{2} - (p+q)^{2} $, which
in other words fixes the normalization point of local operators arising
in an expansion of the non-local matrix elements (5-7). The corresponding
anomalous dimensions can be found in the literature.

The WF's $ g^{(1)} \; , \;  g^{(2)}$ represent twist 4 contributions to
the two-particle photon WF (5). Using equations of motion one can relate
them with three paricle WF's of twist 4 which include an additional
gluon \cite{BBK,BF1}. These three-particle WF's would appear in the
correlation function (3) if a soft gluon emission from the heavy
quark is taken into account. Such a contribution is typically rather small
and will be omitted in what follows. On the other hand, we retain the
two-particle twist 4 contribution due to $ g^{(1)} \; , \;  g^{(2)}$.
That is, the most important twist 4 effects are included in our calculation.
Finally, adding a perturbative photon emission from the u- and b- quarks
we arrive at the following expressions for the invariant functions $
T_{1} \; , \; T_{2} $ :
\bea
T_{1} &=& \int_{0}^{1} \frac{1}{m_{b}^{2} - (p+uq)^{2}} \left[ e_{u} \ll
\bar{u} u \rl \left( \chi \phi(u) - 4 \frac{g^{(1)}(u) -
g^{(2)}(u)}{m_{b}^{2}
 - (p+uq)^{2}} (1 + \frac{2 m_{b}^{2}}{m_{b}^{2}- (p+uq)^{2}})
\right) \right. \nonumber \\
  &+& \left.\frac{m_{b}}{2} f \frac{g_{\perp}(u)}{m_{b}^{2}- (p+uq)^{2}} +
\frac{3 m_{b}}{4 \pi^{2}} \left( (e_{u} - e_{b})
\bar{u} \frac{m_{b}^{2} - p^{2}}{m_{b}^{2} - \bar{u}p^{2}} + e_{b} \ln{
\frac{m_{b}^{2} - \bar{u}p^{2}}{u m_{b}^{2}}} \right) \right] \\
T_{2} &=& \int_{0}^{1} \frac{1}{m_{b}^{2} - (p+uq)^{2}} \left[ e_{u} \ll
\bar{u} u \rl \left( \chi \phi(u) - 4 \frac{g^{(1)}(u)}{m_{b}^{2}  -
(p+uq)^{2}}
(1 + \frac{2 m_{b}^{2}}{m_{b}^{2}- (p+uq)^{2}}) \right) \right. \nonumber \\
      &+& \left. \frac{3 m_{b}^{3}}{4 \pi^{2} (m_{b}^{2} - p^{2})} \left\{
\left(
 (e_{u}- e_{b})( u - \bar{u} + \frac{p^{2}}{m_{b}^{2}} - \frac{p^{2} u^{2}}{
m_{b}^{2} - \bar{u} p^{2}} ) - (e_{u} + e_{b})u \frac{p^{2}}{m_{b}^{2}}
  \right) \frac{\bar{u}(m_{b}^{2} - p^{2})}{m_{b}^{2} - \bar{u} p^{2}}
 \right. \right.  \nonumber \\
&+& \left. \left. e_{b}(u - \bar{u} + \frac{p^{2}}{m_{b}^{2}})
\ln{\frac{m_{b}^{2} - \bar{u}p^{2}}{u m_{b}^{2}}} \right\} \right]
\eea
  We now proceed
to a description of the photon WF's (5-6).

\section{Light cone photon wave functions}

The leading twist photon WF $ \phi (u , \mu^{2}) $ has been
introduced and investigated
in detail in Ref.\cite{BBK} in connection with a study of the radiative
decay $ \Sigma \rightarrow p + \gamma $. Formally, the photon matrix
element of
an arbitrary (non-local) operator is defined as a weighted sum
of matrix elements of the same operator between the vacuum and (transversely
polarized) vector meson states :
\bea
\ll \gamma (q) | \ldots | 0 \rl &=& i \sqrt{4 \pi \alpha} e_{\mu}^{(\lambda)}
\int dz \; e^{iqz} \ll 0 | T { j_{\mu}^{em}(z) \ldots } | 0 \rl \nonumber \\
                                &=& \sqrt{4 \pi \alpha} \sum_{n}
\frac{1}{g_{n}} \ll V_{n}(q)| \ldots | 0 \rl \; ,
\eea
where $ j_{\mu}^{em} = \frac{2}{3} \bar{u} \gmmu u - \frac{1}{3} \bar{d}
\gmmu d - \frac{1}{3} \bar{s} \gmmu s $ and the couplings $ g_{n} $ are
defined via
\beq
\ll 0 | j_{\mu}^{em}| V_{n} \rl = e_{\mu}^{(\lambda)} \frac{m_{n}^{2}}{g_{n}}
\; \; \; , \; \; \; g_{\rho} \simeq 5.5
\eeq
The particular value for $ g_{\rho} $ quoted above has been found long ago
by the QCD sum rules method \cite{SVZ}.

Our definition (10) is equivalent to the one given in \cite{BBK}, \cite{BF1}
for the case of the leading twist photon WF $ \phi_{\perp}(u) $. There it
has been defined as the vacuum expectation value of the non-local
operator $ \bar{q} \sigma_{\mu \nu}q $ (where $ q = u,d,s $ ) in an
external electromagnetic field $ F_{\alpha \beta}(x) $. The QCD action is
modified in this case to $ \int dz \, L_{QCD}(z) + \sqrt{4 \pi \alpha} \int
dz \, A_{\mu}(z) j_{\mu}^{em}(z) $ and, expanding to the first order in
the plane wave field strength $ F_{\alpha \beta}(x) = i
(e_{\beta}^{(\lambda)} q_{\alpha} - e_{\alpha}^{(\lambda)} q_{\beta}) e^{iqx}
$ , we obtain
\beq
\int dz \; e^{iqx} \ll 0 | T \{ j_{\mu}^{em}(z) \bar{q}(x) \sigma_{\alpha
\beta} q(0) \} | 0 \rl = e_{q} \ll \bar{q} q \rl \chi (q_{\beta}
\delta_{\alpha \mu} -  q_{\alpha}\delta_{\beta \mu}) \int du \;  e^{iuqx}
\phi(u)
\eeq
Thus, according to Eq.(10), one gets
\beq
\ll \gamma(q) | \bar{u}(x) \sigma_{\alpha \beta} u(0) | 0 \rl = i \sqrt{4
\pi \alpha} e_{u} \ll \bar{u} u \rl \chi (e_{\alpha}^{(\lambda)}
q_{\beta} - e_{\beta}^{(\lambda)} q_{\alpha}) \int du \;  e^{iuqx}
\phi(u) \; ,
\eeq
i.e. Eq.(5). In the constant external field limit $
q \rightarrow 0 $, Eq.(12) goes back to the definition of a quantity
known as the magnetic susceptibility of the quark condensate $ \chi
$\cite{IS}:
\beq
\ll \bar{q} \sigma_{\alpha \beta} q \rl_{F} = e_{q} \ll \bar{q}
 q \rl \chi F_{\alpha \beta}
\eeq
The magnetic susceptibility $ \chi $ has been found \cite{BK}, \cite{BKo}
by the QCD sum rules method. At an accuracy of the order of 30 \% it is
dominated by the $ \rho $-meson contribution : at $ \mu^{2} \simeq 1 \;
GeV^{2} $
\[
\chi = \left\{ \begin{array}{ll}
            - 4.4  \; Gev^{-2} &  \rho, \rho' , \rho'' \; are \; included \\
            - 3.3  \; Gev^{-2} &  \rho \; is \; included
               \end{array}
       \right.
\]
It has been suggested on these  grounds that the vector dominance works
sufficiently well for electromagnetic properties of hadrons in the long wave
limit $q \rightarrow 0$.
Thus, to the quoted accuracy, by virtue of the local version of Eq.(10)
we obtain the following value for the normalization constant $ f$ in
Eq. (6) :
\beq
f \simeq  \frac{e_{u}}{g_{\rho}} f_{\rho}m_{\rho} \; \; , \; \; f_{\rho}
\simeq 200 \; MeV
\eeq
The validity of Eq.(15) can be checked by the standard  QCD sum
rules technique.

Here some comments are in order. It is known that two-particle
operators containing $ \gmmu $ and $ \gmmu \gmf $  instead of $
\sigma_{\mu \nu} $ have non-zero projections on a vector meson state.
The corresponding wave functions ( call them $ g_{\perp}^{v} $ and $
g_{\perp} $, respectively) contain pieces of different twist. As
has been shown in Ref. \cite{ABS}, a twist 2 contribution to a vector meson
WF  $ g_{\perp}^{v} $ can be expressed via an integral of a longitudinally
polarized vector meson WF of the leading twist 2. A similar relation holds
for the derivative $ d g_{\perp}(u)/ du $ (see Eq.(39) of \cite{ABS} ).
Thus we conclude that in the
case of photon the WF  $ g_{\perp}^{v} $ vanishes to the leading twist
accuracy, while the WF  $ g_{\perp}^{v} $ is a constant which is set to be
unity with
our normalization. At this point we differ from similar calculations in
Refs. \cite{AB,KSW} where a contribution due to $ g_{\perp} $ was omitted.
Numerically, the effect of the inclusion of the corresponding term in the
sum rule (see Eq. (19) below) is about 25 \% for the decay width.

Let us now discuss the much less trivial problem of finding the functional
$u$-dependence of the WF's (5-6). As a non-local operator is
equivalent to an infinite series of local operators , this task might
appear formidable. Indeed, in order to restore the
functional $u$-dependence, one has to know all moments $ \int_{0}^{1} du
\; (2u -1)^{n} f(u) $, i.e. , according to (5-6), all matrix elements of
the type $ \ll \gamma | \bar{q} ( \stackrel{\leftrightarrow}{\nabla}_{\alpha}
x_{\alpha} )^{n} \sigma_{\mu \nu} q | 0 \rl $. The way out lies in the use
of an approximate conformal symmetry of QCD which holds at the one-loop
level.
The conformal invariance permits an expansion of a non-local operator of
the type (5-6) in a series over multiplicatively renormalizable operators
with ordered anomalous dimensions, which induces the expansion of the leading
twist WF in the series of Gegenbauer polynomials
\cite{BBK},\cite{BF1},\cite{BF2}.
The asymptotic WF's are defined as contributions of
operators with the lowest conformal spin and unambiguously fixed by the
group structure. Pre-asymptotic corrections correspond to the operators
with the next-to-leading conformal spin, whose numerical values are usually
calculated by the QCD sum rules method.
It was found \cite{BBK} that for the leading twist photon WF
contributions of higher conformal spins are small. In terms of the
dispersion relation this means that the corrections to the asymptotic WF's
of the $ \rho, \rho', \rho'' $ states have opposite signs and nearly
cancel in the sum.
Thus, the simple asymptotic formula holds (in what follows, we use $ \bar{u}
\equiv 1-u $)
\beq
\phi_{\perp}(u) = 6 u \bar{u}
\eeq
To the leading twist accuracy we use for the WF (6) $ g_{\perp}(u) = 1 $
. For the twist-4 WF's $ g^{(1)}(u) \;, \; g^{(2)}(u) $ we use the
following expressions (see \cite{AB}):
\bea
g^{(1)}(u)                 &=& -\frac{1}{8}\bar{u}(3-u) \nonumber \\
g^{(2)}(u)                 &=& -\frac{1}{4} \bar{u}^{2}
\eea

\section{Form factors and the decay width}

To match the answers (8,9) with the B-meson contribution to the correlation
function (3), we note that (8,9) can be re-written as the dispersion
integrals with the expression $ (m_{b}^{2} - \bar{u} p^{2})/u $ being the
mass of the intermediate state. The duality prescription requires that
this invariant mass has to be restricted from above by the duality threshold
$ s_{0} \simeq 35 \; GeV^{2} $ (this value is obtained from corresponding
two-point sum rules). As it is easy to see, this transforms into an
effective cut-off in the lower limit of the u-integral \cite{BKR},\cite{BH}.
Finally,
we make the standard Borel transformation suppressing both higher states
resonances and higher Fock states in the full photon wave function.
Under the Borel transformation $ -(p+q)^{2} \rightarrow M^{2} $
\bea
\frac{1}{m_{B}^{2} - (p+q)^{2}} \rightarrow \exp{(-
\frac{m_{B}^{2}}{M^{2}}) } \nonumber \\
\frac{1}{ m_{b}^{2} - ( p+uq)^{2}} \rightarrow \frac{1}{u} \exp{( -
\frac{m_{b}^{2} - \bar{u}p^{2}}{ u M^{2}})}
\eea
Our final sum rules take the form
\bea
g(p^{2}) &=&  \frac{m_{b}}{ f_{B}}
\int_{0}^{1} \frac{du}{u}
 \; \exp{( \frac{m_{B}^{2}}{M^{2}} - \frac{m_{b}^{2} - \bar{u}
p^{2}}{u
M^{2}}) } \Theta(u - \frac{m_{b}^{2} - p^{2}}{ s_{0} - p^{2}})
\times \nonumber \\
& & \left[  e_{u} \ll \bar{u} u \rl \left( \chi \phi(u)
-  4 (g^{(1)}(u) - g^{(2)}(u)) \frac{m_{b}^{2} +u
M^{2}}{u^{2}M^{4}} \right) + \frac{m_{b}f}{2u M^{2}} g_{\perp}(u) \right.
\nonumber  \\
&+& \left. \frac{3m_{b}}{4 \pi^{2}} \left\{ (e_{u} - e_{b})\bar{u}
\frac{m_{b}^{2} - p^{2}}{m_{b}^{2} - \bar{u}p^{2}} + e_{b} \ln{
\frac{m_{b}^{2} - \bar{u}p^{2}}{u m_{b}^{2}}} \right\} \right]  \\
f(p^{2}) &=&  \frac{m_{b}}{f_{B}}
\int_{0}^{1} \frac{du}{u}
 \; \exp{( \frac{m_{B}^{2}}{M^{2}} - \frac{m_{b}^{2} - \bar{u}
p^{2}}{u
M^{2}}) } \Theta(u - \frac{m_{b}^{2} - p^{2}}{ s_{0} - p^{2}})
\times \nonumber \\
& & \left[  e_{u} \ll \bar{u} u \rl \left( \chi \phi(u)
-  4 g^{(1)}(u)\frac{m_{b}^{2} +u
M^{2}}{u^{2}M^{4}} \right) \right. \nonumber \\
&+& \left.\frac{3 m_{b}^{3}}{4 \pi^{2} (m_{b}^{2} - p^{2})}
\left\{ \left( (e_{u}- e_{b})( u - \bar{u} + \frac{p^{2}}{m_{b}^{2}} -
\frac{p^{2} u^{2}}{m_{b}^{2} - \bar{u} p^{2}} )  \right. \right. \right.
\nonumber \\
&-& \left. \left. \left. (e_{u} + e_{b})u \frac{p^{2}}{m_{b}^{2}} \right)
\frac{\bar{u}(m_{b}^{2} - p^{2})}{m_{b}^{2} - \bar{u} p^{2}}
+  e_{b}(u - \bar{u} + \frac{p^{2}}{m_{b}^{2}})
\ln{\frac{m_{b}^{2} - \bar{u}p^{2}}{u m_{b}^{2}}} \right\} \right]
\eea
We now turn to numerical estimates. In evaluating (19,20), we have
used
the following set of parameters : $ m_{b} = 4.7 \; GeV , \; m_{B} = 5.28
\; GeV , \; s_{0} \simeq 35 \; GeV^{2} , \; f_{B} \simeq 135 \; MeV $
\cite{BKR},\cite{BBKR}. The particular value of $ f_{B} $ corresponds
to the result of the sum rule calculation with $ O( \alpha_{s}
) $ corrections omitted, cf. \cite{BKR}, \cite{BBKR}. Within our accuracy we
also neglect the logarithmic
evolution of the WF's to a higher normalization point in the sum rules
(19,20) which is of order of the characteristic virtuality of  the heavy
quark
in the B-meson.
The Borel mass $ M^{2} $ has been varied in
the interval from 8 to 20 $ GeV^{2} $. We have found that within the
variation of $M^{2}$ in this region, the result changes by less than
10 \%. As the sum rules of the type (19,20) are expected to work in the
region $ m_{b}^{2} - p^{2} \sim $ a few $GeV^{2} $, which is somewhat smaller
than the maximal available $ p^{2} = m_{b}^{2} $, we have to use some
extrapolation formulas to extent the results of the sum rules to the whole
region of $  p^{2} $. We have found that the best agreement is reached
with the dipole formulas
\beq
g(p^{2})  \simeq \frac{h_{1}}{(1 -
\frac{p^{2}}{m_{1}^{2}})^{2}}
\; \; \;, \; \; \;  f(p^{2}) \simeq \frac{h_{2}}{(1 -
\frac{p^{2}}{m_{2}^{2}})^{2}}
\eeq
The fact that the dipole approximation agrees better with results of the
sum rules at not too large $ p^{2} $ than the
pole-type formulas has been also noted in \cite{ABS} for analogous sum
rules including vector mesons. We find
\bea
h_{1} \simeq 1 \; GeV \; \; \; , \; \; \; m_{1} \simeq 5.6 \; GeV
\nonumber \\
h_{2} \simeq 0.8 \; GeV \; \; \; , \; \; \; m_{2} \simeq 6.5 \; GeV
\eea
   The calculation of the decay width yields
\beq
\Gamma = \frac{ \alpha G_{F}^{2} |V_{ub}|^{2} m_{B}^{5}}{ 96 \pi^{2}} \, I
\eeq
where ($ \bar{x} = 1-x $)
\beq
m_{B}^{2} I = \int_{0}^{1} dx \, \bar{x}^{3}x \{ g^{2}(x) + f^{2}(x) \}
\eeq
Here $ x = 1 - 2 E/m_{B} $ and $ E $ stands for the photon energy.
The corresponding differential width can be read off Eqs. (23,24). We
expect the accuracy for $ I $ to be of the order of 20-30 \%. The
numerical answer for the quantity $ I $ is $ I \simeq 0.021 $ \\
( $ I
\simeq 0.015 $ if the contribution due to  $ g_{\perp} $ is omitted).

We would like now to compare our result with that obtained in Ref.
\cite{AES} within the non-relativistic quark model. These authors have
obtained the following answer for the decay width :
\beq
\Gamma = \frac{\alpha G_{F}^{2} |V_{ub}|^{2} m_{B}^{5}}{48 \pi^{2}}
\frac{Q_{u}^{2}}{6} \frac{f_{B}^{2}}{m_{u}^{2}} \; ,
\eeq
where now $ f_{B} $ stands for the physical value $ f_{B} \simeq 175 \;
MeV $ and a small contribution stemming from photon emission off the
b-quark has been neglected. Note that in Eq. (23) the constant $ f_{B} $
appears implicitly
in the denominator (see Eqs. (19,20)), while it is in the numerator of
Eq.(25). This
difference stems from the use of the dispersion approach. We find that
Eq.(25) yields a result numerically close to (23) with the choice of
the free parameter
\beq
m_{u} \simeq 480 \; \; MeV \; ,
\eeq
The photon spectra corresponding to the light cone QCD calculation and
the quark model approach are compared in Fig.1 using the same
normalization. One sees that while the quark model gives a fully
symmetric spectrum, in our approach the spectrum is
slightly asymmetric, as a result of a balance between a typical highly
asymmetric  resonance-type behaviour given by the non-perturbative
contributions, and a perturbative photon emission.
Both curves emphasize
the dominant role of hard photons in the decay of interest and vanish at
the end-points.

Finally, we present the estimated rate for the decay of interest
\beq
Br( B \rightarrow l \nu \gamma ) \simeq 2 \cdot 10^{-6} \; \; ,
\eeq
which corresponds to the choice $ m_{u} = 480 \; MeV $. We have used
here the values $ |V_{ub}|/ |V_{cb}| = 0.08 $ , $ |V_{cb}| = 0.04 $
\cite{PDG}.

\section{Summary}

We have calculated within the light cone QCD sum rules technique the
transition form factors and the differential and total width of the
radiative decay $ B \rightarrow l \nu \gamma $. In accordance with the
expectations of Ref. \cite{AES}, the photon emission overcomes the
helicity suppression of the pure leptonic $ B \rightarrow l \nu $
amplitude and yields an experimentally admissible decay  width. Comparing
our result with that suggested by the quark model, we have fixed the
value of the effective constituent quark mass $ m_{u} \simeq 480 \;
MeV $, which is consistent with estimates for the "inertia parameter"
$ \bar{\Lambda}_{u} $ in the heavy quark effective theory. The photon spectra
calculated within our approach and the quark model calculation
have similar shapes. They both emphasize hard photon
emission, vanish at the boundaries of the spectrum
and do not exhibit the typical bremsstrahlung  $ 1/E $ behavior.
The completely symmetric form given by the quark model \cite{AES}
is an artifact of the absence of hadronic poles. The light cone QCD
approach gives a slightly asymmetric spectrum due to a balance between
resonance-type non-perturbative contributions and perturbative
photon emission.

In the perturbative QCD approach to exclusive processes \cite{CZ}
annihilation amplitudes are typically expressed via a convolution of wave
functions of participating hadrons. These integrals turn out to be
logarithmically divergent in the end-point region \cite{CZ}, which
indicates the breakdown of perturbative factorization and suggests that
the annihilation mechanism is of long distance type. This
phenomenon is a manifestation of a long distance dominance, akin to an
appearence of the $ 1/m_{q} $ factor in the quark model approach.
In this sense the situation is similar to the problem of the Feynman-like
end-point contribution to the pion form factor \cite{BH}. We believe that
the dispersion light cone QCD approach, suggested in this context in Ref.
\cite{BH} and used in the present letter for the particular annihilation-like
process, may be of use also for reactions involving gluon annihilation.

We conclude by remarking that this process may be
useful to extact the KM matrix element $ |V_{ub}| $ , since the  value
obtained for the branching
ratio $ Br( B \rightarrow \mu \nu \gamma ) \simeq 2 \cdot 10^{-6} $ is
reasonably large and, in addition, is theoretically under control in view
of the simplicity of the hadronic system.

{\it Note added} \\
After completion of the previous version of this work we became aware of
recent papers \cite{AB,KSW}
where a similar technique has been applied to the radiative decays $ B
\rightarrow \rho + \gamma $ and $ B \rightarrow \omega + \gamma $ and in
\cite{KSW} to the $ B \rightarrow l \nu \gamma $ decay.

 {\it Acknowledgement} \\

The research of G.E. was supported in part by GIF
and by the Fund for the Promotion of Research at the Technion. The work
of R.R.M. was supported in part by the Natural Sciences and Engineering
Research Council of Canada. We are grateful to V.M. Braun for pointing
out errors in an earlier version of this work where we used an incorrect
set of photon wave functions of leading twist and did not take into account
perturbative photon emission.

\clearpage
\begin{figure}
\caption[xxx]{The normalized differential width $ \Delta \equiv 1/ \Gamma
(d \Gamma / d x ) $ of the
decay $ B \rightarrow l \nu \gamma $ as functions of $ x = 1 - 2 E /m_{B}$
where E is the photon energy.  The solid and dashed curves correspond to
the results of the light cone sum rule calculation (Eqs.(25,26)) with the
dipole ansatz for the form factor $ f(p^{2}) $ and the quark model
calculation of Ref. \cite{AES}, respectively. }
\end{figure}


\clearpage
\begin{thebibliography}{99}

\bibitem{BSS}
M.~Bander, D.~Silverman and A.~Soni, Phys. Rev. Lett. {\bf 44} (1980) 7:
                                     (E) {\bf 44} (1980) 962.

\bibitem{EMM}
G.~Eilam, B.~Margolis and R.~Mendel, Phys. Lett. {\bf B185} (1986) 145.

\bibitem{DHT}
N.G.~Deshpande, X.-G.~He and J.Tranpetic, OITS-564-REV, hep-ph/9412222.

\bibitem{MS}
R.R.~Mendel and  P.~Sitarski, Phys. Rev. {\bf D36} (1987) 953.

\bibitem{BGW}
G.~Burdman, T.~Goldman and D.~Wyler, Phys. Rev. {\bf D51} (1995)
111.

\bibitem{ABBS}
D.~Atwood, B.~Blok and A.~Soni, SLAC-PUB-6635, TECHNION-PH-94-11,
hep-ph/9408373.

\bibitem{AES}
D.~Atwood, G.~Eilam and A.~Soni, SLAC-PUB-6716, TECHNION-PH-94-13,
hep-ph/9411367.

\bibitem{EJM}
G.~Eilam, A.~Ioannissian and R.R.~Mendel, TECHNION-PH-95-4, hep-ph/9505222.

\bibitem{Neu}
M.~Neubert, Phys. Rep. {\bf 245} (1994) 259.

\bibitem{SVZ}
M.I.~Shifman, A.I.~Vainshtein and V.I.~Zakharov, Nucl. Phys. {\bf B147}
(1979) 385.

\bibitem{BBK}
I.I.~Balitsky, V.M.~Braun and A.V.Kolesnichenko, Nucl. Phys. {\bf B312}
(1989) 509.

\bibitem{BF1}
V.M.~Braun and I.E.~Filyanov ,
Z. Phys. {\bf C44} (1989) 157.

\bibitem{BH}
V.~Braun and I.~Halperin, Phys. Lett. {\bf B328} (1994) 457.

\bibitem{BBKR}
V.~Belyaev, V.~Braun, A.~Khodjamirian and R.~Ruckl, MPI-PhT/94-62,
CEBAF-TH-94-22, LMU 15/94, hep-ph/9410280.

\bibitem{BKR}
V.M.~Belyaev, A.Khodjamirian and R.Ruckl, Z. Phys. {\bf C60} (1993) 349.

\bibitem{ABS}
A.~Ali, V.~Braun and H.Simma, Z. Phys. {\bf C63} (1994) 437.

\bibitem{CZ}
V.L.~Chernyak and A.R.~Zhitnitsky, Phys. Rep. {\bf 112} (1984) 173.

\bibitem{BF2}
V.M.~Braun and I.E.~Filyanov, Z. Phys. {\bf C48} (1990) 239.

\bibitem{IS}
B.L.~Ioffe and A.V.~Smilga, Nucl. Phys. {\bf B216} (1983) 373.

\bibitem{BK}
V.M.~Belyaev and Ya.I.~Kogan, Yad. Fiz. {\bf 40} (1984) 1035.

\bibitem{BKo}
I.I.~Balitsly and A.V.~Kolesnichenko, Yad. Fiz. {\bf 41} (1985) 282.

\bibitem{PDG}
Particle Data Group, Phys. Rev. {\bf D50} (1994) 1173.

\bibitem{AB}
A.~Ali and V.M.~Braun, DESY 95-106, hep-ph/9506248.

\bibitem{KSW}
A.~Khodjamirian, G.~Stoll and D.~Wyler, ZU-TH 8/95, LMU-06/95,
hep-ph/9506242.

 \end{thebibliography}
\end{document}